\documentclass[10pt]{article}
\usepackage[OE]{express}

\usepackage{amsmath,amssymb}
\usepackage{verbatim}
\usepackage{float}

\DeclareMathOperator{\Tr}{Tr}

\usepackage[utf8]{inputenc}
\usepackage{color}
\usepackage{graphicx}

\begin{document}
\title{Pulsed quantum continuous-variable optoelectromechanical transducer}

\author{Nikita Vostrosablin,\authormark{1*} Andrey A. Rakhubovsky,\authormark{1} and Radim Filip\authormark{1}}

\address{\authormark{1}Department of Optics, Palack{\'y} University, 17. Listopadu 12, 771 46 Olomouc, Czech Republic}

\email{\authormark{*}nikita.vostrosablin@upol.cz} 

\newcommand{\AndreyRNote}[1]{\textsf{[\textbf{AR}: #1]}}
\newcommand{\ar}[1]{\AndreyRNote{#1}\marginpar{\textcolor{red}{$\bigstar$}}}



\begin{abstract}
We propose a setup allowing to entangle two directly non-interacting radiation modes applying four sequential pulsed quantum resonant interactions with a noisy vibrational mode of a mechanical oscillator which plays the role of the mediator. We analyze Gaussian entanglement of the radiation modes generated by the transducer and confirm that the noisy mechanical mode can mediate generation of entanglement. The entanglement, however, is limited if the interaction gains are not individually optimized.  We prove the robustness of the transducer to optical losses and the influence of the mechanical bath and propose the ways to achieve maximal performance through the individual optimization.
\end{abstract}



\section{Introduction}

Quantum transducers are hybrid quantum systems important for development
of unified quantum technology~\cite{Kimble2008}. They practically demonstrate ability to
universally entangle even very different quantum systems~\cite{Verdu2009,Marcos2010, Hafezi2012} and therefore,
exchange quantum states between them. The
transducer in principle connects two different systems $A$ and $B$ that otherwise are not
interacting~\cite{Regal2011}. For an example, the systems $A$ and $B$ individually interact only in the pairs $A-M$ and
$B-M$ with a mediating system $M$. The latter is however also a quantum system,
therefore it can introduce quantum noise to the transducer. Moreover, $M$ is
typically open to an environment, which is noisy and lossy and limitedly
measurable. It is therefore important to take this connection into account to propose
a feasible quantum transducer. Continuous-variable~(CV) quantum transducers are
capable to quantum mechanically couple two different oscillators $A$ and
$B$ by a mediating oscillator $M$. They can generate Gaussian CV entanglement, which can be used, for example, to teleport states between $A$ and $B$~\cite{Takeda2013}.  Advantageously, they can be built, without any nonlinearity, from
the most common linearized interactions $A-M$, $B-M$ of the oscillators.
First type of transducers use simultaneously running linearized couplings $A-M$ and
$B-M$ towards a steady state where $A-B$ coupling can be of a high quality
and sufficient strength~\cite{LinTian2015}. Ideally, the
mediator $M$ should be completely eliminated and not influence the coupling $A-B$. Nontrivial optimization of the $A-M$ and $B-M$ coupling strengths
over time can improve the transducer quality.

These transducers, however, cannot operate in time-resolved quantum regime,
with nonclassical states defined within a shorter time interval, used in
modern optical~\cite{Ogawa2016, Makino2016} and
microwave experiments~\cite{Palomaki2013, Lehnert2013}. To solve this problem, pulsed CV quantum
transducers operating with optical and microwave pulses are required. The pulsed regime was already used to generate entanglement~\cite{Palomaki2013} and propose for quantum teleportation~\cite{Hofer2011}.
Complementary to previous approach, the pulsed transducers individually control the
interactions $A-M$ and $B-M$ by time non-overlapping pump pulses. The main
idea is to use twice a sequence of the interactions $A-M$ and $B-M$ and exploit power of
geometric phase effect~\cite{Berry1987}  for CVs to eliminate the mediator $M$ regardless of its noisy initial state.
Recently, a principal robustness of such the pulsed transducers has been
proven theoretically and temporal optimization of pulse control beyond the geometric phase effect has been suggested to
reach the robust regime~\cite{Kupcik2015}. It opens a way to propose the pulsed transducers
for various experimental platforms, for example, quantum optomechanics and electromechanics.

Rapid development of quantum optomechanics~\cite{Aspelmeyer2014, Metcalfe2014}  puts forward a mechanical oscillator as a suitable mediator for the construction of the pulsed transducers. A lot of progress is done in the direction of optoelectromechanical transducers. The reversible  optoelectromechanical transducer which used the effective source of two-mode squeezing between optical idler and a microwave signal to transfer quantum states between optical and microwave fields via quantum teleportation was proposed in~\cite{Barzanjeh2012}. High-fidelity quantum state conversion between microwave and optical fields may be performed through the excitation  of the mechanical dark mode~\cite{Regal2011, Wang2012, Lin2012, YDWang2012, McGee2013}. This approach allows significant suppression of the mechanical noise from the mediator. Several experimental works were performed in the direction of quantum state transfer. In~\cite{Winger2011} an integrated optomechanical and electromechanical nanocavity was used to efficiently interconvert microwave and optical signals. In this device a photonic crystal defect cavity and an electrical circuit were both coupled to the same mechanical degree of freedom. In~\cite{Bochmann2013} a piezoelectric optomechanical crystal was used for coherent signal transfer between itinerant microwave and optical fields. In another experimental work~\cite{Andrews2014} a mechanically compliant silicon nitride membrane was used to realize a high-fidelity conversion between optical light and microwave. In~\cite{Bagci2014} a transducer utilizing a high-Q nanomembrane to interconvert radio-frequency waves with optical light was demonstrated. Very recently a new experimental work~\cite{Menke2017} considering the device capable for microwave-to-optics conversion by placing all components inside a re-entrant microwave cavity was performed. This design allows the wireless coupling to the transmission line with the possibility to vary the strength of this coupling without affecting the performance of the setup. There was also sufficient progress in the domain of transducers interconnecting optical fields. Optomechanical device entangling two optical fields was proposed in~\cite{Asjad2015}. This setup was also considered to be used for quantum state teleportation of light signals over long distances, mediated by concatenated swap operations. The scheme containing two-mode optical cavity and the closed-loop feedback control was studied in~\cite{Asjad2016}. This setup allowed to entangle the outputs and to coherently teleport quantum states between them. Coherent quantum state transfer between optical fields by the sequence of optomechanical pulses was studied theoretically~\cite{Tian2010}. In the experiment~\cite{Dong2015} a conversion of optical fields between two different frequencies by coupling them to a mechanical mode of a silica resonator was demonstrated. It was also the first experimental observation of a mechanical dark mode for the optomechanical transducer. Despite this remarkable progress in the field of opto- and electromechanical transducers, there is still one sufficient limitation~--- mechanical noise which restricts the performance of aforementioned setups. Our approach, based on the geometric phase effect allows to bypass this limitation. This method requires a pulsed control of mechanical systems, which is simultaneously advantageously compatible with modern quantum optics~\cite{Ogawa2016, Makino2016, Miwa2014}.

In this paper, we propose a pulsed CV quantum
transducer with a noisy mechanical system as a mediator and analyze its feasibility for optoelectromechanical experiment. We study the influence of radiation losses and mechanical bath and we show that optimization of parameters allows high performance of the proposed transducer even for very noisy mediator. To demonstrate the feasibility of the proposed setup we firstly consider the symmetrical transducer which connects optical field to optical field. Such scheme is a good demonstration of the viability of the proposed concept and does not require involved modifications of the state-of-the-art experimental platforms for near-future implementation. Only then we consider more general case of the asymmetrical transducer coupling optical to microwave fields. This case is very important since it follows the trend to connect different quantum systems which is crucial for the future development of hybrid quantum systems~\cite{Aspelmeyer_book2014,Wallquist2009}.

\section{Pulsed CV Quantum Symmetrical Transducer}
\label{perform}

\subsection{Setup description}
\label{setup_descr}

\begin{figure}
\centering \includegraphics[width=0.7\linewidth]{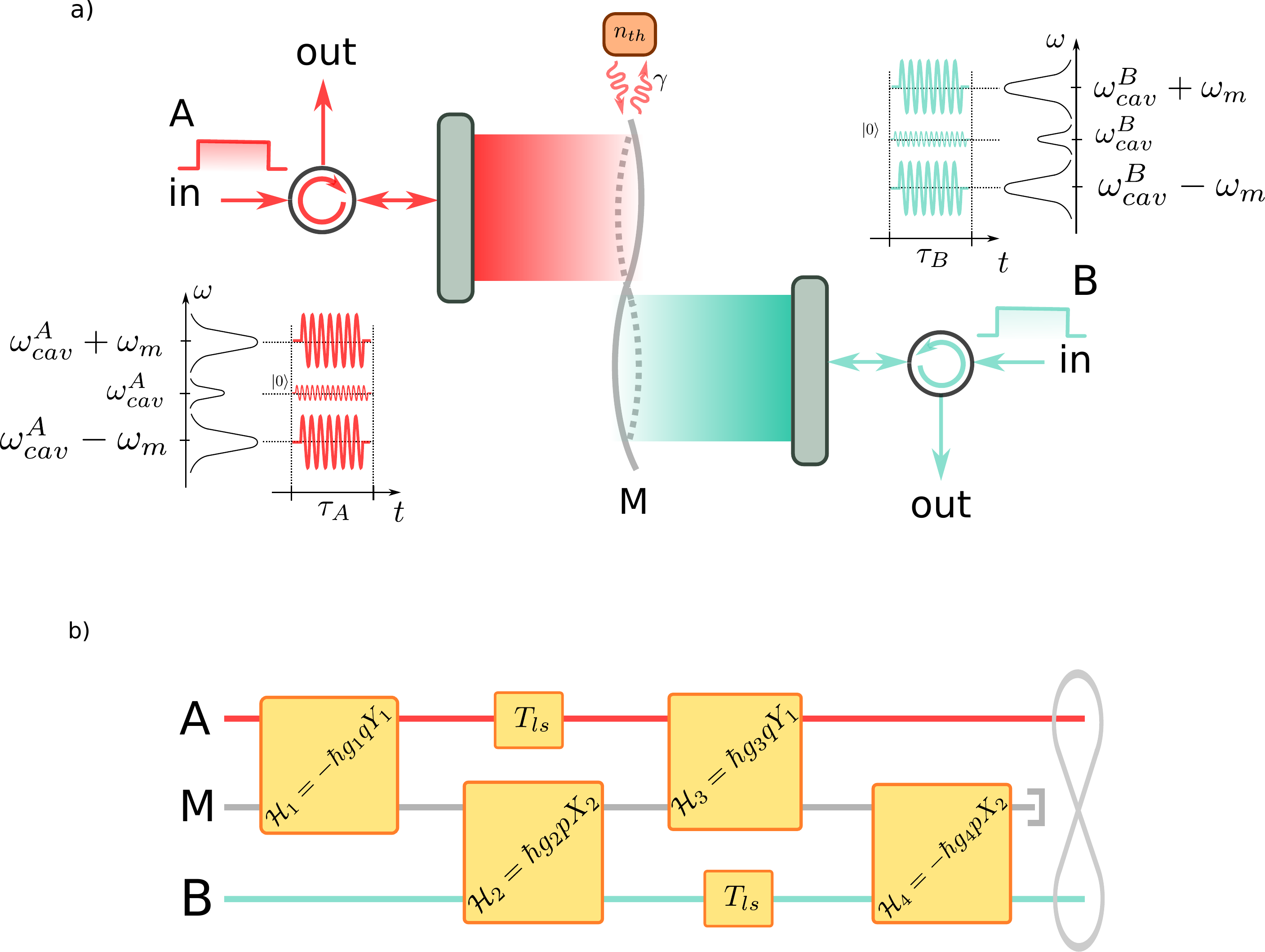}
\caption{\label{fig1}(a)~Schematic representation of the pulsed transducer interconnecting two radiation modes $A$ and $B$ which may be both optical as well as optical and microwave fields. The first amplitude-modulated pulse $A$ (red) containing classical Stokes and anti-Stokes sidebands on $\omega^A_{cav} \pm \omega_m$ and quantum fluctuations, which are in the vacuum state $|0 \rangle$, on $\omega^A_{cav}$, is sent to the first cavity to interact with mechanical mode $M$ during the time $\tau_A$ . After the interaction is complete the pulse is sent to the delay line whereas the second pulse $B$ (blue), being in the vacuum state $|0 \rangle$ as well, interacts with $M$ during $\tau_B$ within the second cavity. The interactions of the two pulses are then repeated one more time and both pulses are released to the outputs. Radiation pulses are subject to losses $T_{ls}$ and the mechanical mode is coupled at rate $\gamma$ to the mechanical bath with mean occupation number $n_{th}$. (b)~Block-diagram of the sequence of the interactions between the pulses of modes $A$ and $B$ with the mechanical mediator $M$.}
\end{figure}

The basic idea of the setup which we consider is depicted in Fig.~\ref{fig1}.  The two radiation modes $A$ and $B$ are coupled to the same mechanical oscillator $M$ but do not interact directly.  The quantum states of the modes $A$ and $B$ are defined in temporal pulses with duration $\tau_A$ and $\tau_B$ correspondingly (see Fig.~\ref{fig1}~(a)).  Each pulse interacts to the mechanical mediator $M$ twice during the protocol.  After the interaction of the first pulse with the mechanical mode is complete, the former is sent to the delay line (see Fig.~\ref{fig1}~(a)) while the second pulse enters its cavity to interact with the mechanical mode.  The operation is then repeated until the four interactions are performed.  The block scheme of the protocol is depicted on Fig.~\ref{fig1}~(b).  Note that the transducer between two optical modes requires only one one-sided optomechanical cavity, to which the pulses are directed in turns.

The mechanical mediator is coupled to the pulses by the means of four sequential quantum non-demolition (QND) interactions~\cite{braginsky_quantum_1980,clerk_back-action_2008}. The nondemolition interactions preserve always one quadrature variable (generalized position or momentum) of the single oscillator and perform therefore a partial quadrature exchange between the two oscillators. Only a single non-demolition variable is transferred between two systems during an individual interaction.  The appropriate combination of such QND interactions of the modes $A$ and $B$ with mechanical mediator $M$ allows
driving the latter around a closed path in the phase space in such a way that the geometric phase appears. The geometric phase effect has been already used in quantum optics~\cite{Sorensen2000, Milburn2000, Leibfried2003} and optomechanics where mechanical oscillator was coupled to a qubit~\cite{Vacanti2012} or light~\cite{Pikovsky2012, Khosla2013}.  As a result of the geometric phase imparted to the mechanical system the modes $A$ and $B$ appear to be coupled to each other but not to the mediator $M$ that is brought back to the initial state. This result is achievable due to specific character of the QND interaction, which qualify it to be a basic CV quantum gate. The transducer can therefore principally work for any initial state, even a very noisy one, of the mediator $M$.

In~\cite{Kupcik2015}, it has been observed that the transducer can be stable against the small in-coupling and out-coupling losses of the radiation modes and losses in the delay lines, if the interaction gains of all four QND coupling are optimized. To reach sufficiently high gain of individual interactions with mediator and overall gain of the transducer, the enhancement by a high-$Q$ cavity is necessary. The intracavity field is continuously leaking out the cavity. Simultaneously, the mechanical mode is also continuously damped to its noisy environment. Considerations of these imperfections go far beyond the basic stability check in~\cite{Kupcik2015}. In more realistic setup with the cavities and noisy mechanical environment, we therefore need to carefully analyze the performance of quantum transducer through noisy mediator and compare it to realistic parameters of the experimental schemes.

\subsection{Optomechanical interaction}
\label{opt_inter}

In a basic case an optomechanical system may be modeled as a single cavity mode of the optical resonator interacting with a single one of a mechanical oscillator via the radiation pressure (see Fig.~\ref{fig1}~(a)). The Hamiltonian of the optomechanical system may thus be written as~\cite{law_interaction_1995}:
\begin{equation*}
\label{opt_Ham}
\mathcal H = \frac{\hbar \omega_c}{4} \left( X^2 + Y^2 \right) + \frac{\hbar \omega_m}{4} \left( p^2 + q^2 \right) - \frac{\hbar g_0}{4} q \left( X^2 + Y^2 \right),
\end{equation*}
with $X,Y$ and $q,p$ being the quadratures of optical and mechanical modes correspondingly, with eigenfrequencies  $\omega_c$ and $\omega_m$. These quadratures satisfy commutation relations $[X, Y] = 2 i$, $[q,p] = 2 i$. The single-photon coupling rate~$g_0$ is usually very small and thus the optomechanical interaction is very weak. To further enhance this interaction the cavity is pumped by a strong classical field. This approach allows to linearize the dynamics of the system and consider small quantum corrections to the mean classical values of the quadratures. The CV transducer is capable to generate Gaussian entanglement correlating these corrections.

To obtain a QND interaction we consider each pump to be resonant with the cavity and properly modulated at the mechanical resonant frequency which is assumed to exceed the corresponding cavity decay rates $\omega_m \gg \kappa_A, \kappa_B$ (the resolved sideband condition). After using a rotating wave approximation where we get rid of terms oscillating at $2 \omega_m$, in terms of the quantum corrections defined at linearization, we obtain the following QND interaction Hamiltonian depending on the phase of the pump:
\begin{equation}
\label{QNDH1_H2}
\mathcal H_{i} = \hbar \varkappa_i q Y_A \quad \text{or} \quad \mathcal H_{j} = \hbar \varkappa_j p X_B,
\end{equation}
where $i = \{1,3\}$, $j = \{2,4\}$ denotes the interaction number, $\varkappa_1 = - g_1$, $\varkappa_2 = g_2$, $\varkappa_3 = g_3$, $\varkappa_4 = - g_4$ are individual interaction strengths of radiation modes with the mechanical one.  The change of the sign of interaction strength can be obtained by a suitable adjustment of pump phase. The large intracavity photon number $n_{cav,i}$ corresponding to $i$-th interaction enhances the optomechanical coupling strength so that $g_i = g_0 \sqrt{ n_{cav,i}}$. See Fig.~\ref{fig1}~(b) for our choice of the sequence of the QND interactions.  This sequence of interactions leads to the closed rectangular path in the phase space of mediator's variables $q$ and $p$.  Due to this the mediator becomes uncoupled from the radiation modes at the end of the protocol and does not affect its efficiency.

In the Heisenberg picture the system of quantum Langevin equations~\cite{Giovannetti2001} describing the dynamics of the first and second QND interactions may be written as follows:
\begin{equation}
\label{QNDLang}
\begin{aligned}[l]
& \dot{X}_A = - \kappa_A X_A + \sqrt{2 \kappa_A} X^{in}_A + \varkappa_1 q, \\
& \dot{Y}_A = - \kappa_A Y_A + \sqrt{2 \kappa_A} Y^{in}_A, \\
& \dot{q} = - \frac{\gamma}{2}q + \sqrt{\gamma}\xi_{x1},  \\
& \dot{p} = -\frac{\gamma}{2} p + \sqrt{\gamma} \xi_{p1} - \varkappa_1 Y_A,
\end{aligned}
\qquad
\begin{aligned}[l]
 & \dot{X}_B = - \kappa_B X_2 + \sqrt{2 \kappa_B} X^{in}_B, \\
 & \dot{Y}_B =  - \kappa_B Y_2 + \sqrt{2 \kappa_B} Y^{in}_B - \varkappa_2 p,  \\
 & \dot{q} =  - \frac{\gamma}{2}q + \sqrt{\gamma}\xi_{x2} + \varkappa_2 X_B,  \\
 & \dot{p} = -\frac{\gamma}{2} p + \sqrt{\gamma} \xi_{p2},
\end{aligned}
\end{equation}
where $\kappa_{A,B}$ are cavity decay rates of two corresponding cavities, $\gamma$ is the mechanical damping coefficient and $\xi_{xi,pi}$ are mechanical noise quadratures. Note, here the mechanical decoherence is present during whole the time of the entangling process, differently to simplified analysis in Ref.~\cite{Kupcik2015}.

\subsection{Adiabatic elimination and the entanglement generation}
\label{simple_sect}

As it was mentioned previously we firstly consider symmetrical transducer putting equal decay rates $\kappa_{A} = \kappa_B = \kappa$ and assuming that $\kappa$ is much larger than other  rates in the dynamical equations~\eqref{QNDLang}. The latter condition gives us a possibility to adiabatically eliminate the influence of the intracavtiy field by setting the derivatives of field quadratures equal to zero~\cite{Hofer2011}.  To find theoretical upper bound for generated entanglement, we assume here the mechanical mode decoherence-free putting $\gamma = 0$ and $\xi_{xi, pi} = 0$.  Previous studies~\cite{rakhubovsky_squeezer-based_2016} show that the optomechanical QND interaction can be degraded by cavity memory effects due to finite linewidth $\kappa$ and mechanical bath.  The consideration when these effects are eliminated therefore allows to estimate the ultimate performance of our transducer that we will later use to evaluate the realistic regimes.  We refer to this adiabatic lossless and noiseless regime as the ideal one.

Using Langevin equations and the input-output relations in the form
\begin{equation}
	Q^{out} (t) = \sqrt{2 \kappa} Q(t) - Q^{in} (t),
\end{equation}
where $Q = \{ X, Y \}$, we can show that the scheme depicted in Fig.~\ref{fig1} is equivalent to  the QND interaction between modes $A$ and $B$:
\begin{equation}
\label{qndAE}
\begin{aligned}
  \mathcal X_A^{out} & = \mathcal X_A^{in} + \eta^2 \mathcal X_B^{in},
& \mathcal X_B^{out} & = \mathcal X_B^{in},\\
  \mathcal Y_A^{out} & = \mathcal Y_A^{in},
& \mathcal Y_B^{out} & = \mathcal Y_B^{in} - \eta^2 \mathcal Y_A^{in},
\end{aligned}
\quad
\begin{aligned}
& q = q(0),\\
& p = p(0),
\end{aligned}
\end{equation}
where we have introduced  effective QND coupling strength $\eta = g\sqrt{\frac{2 \tau}{\kappa}}$, new quadratures $\mathcal Q = \{ \mathcal X, \mathcal Y \}$, $\mathcal Q = \frac{1}{\sqrt{\tau}}\int_o^{\tau} Q(t) dt$ integrated over rectangular pulses, pulse duration time $\tau$ (at this point we assumed identical pulses $\tau_A = \tau_B = \tau$) and we have put all optomechanical couplings equal to each other and equal to $g$. As we can see from  \eqref{qndAE} the mechanical mode is completely traced out from these transformations due to geometric phase effect discussed in the Section~\ref{setup_descr} and \ref{opt_inter}.

\begin{figure}[H]
\centering \includegraphics[width=0.7\linewidth]{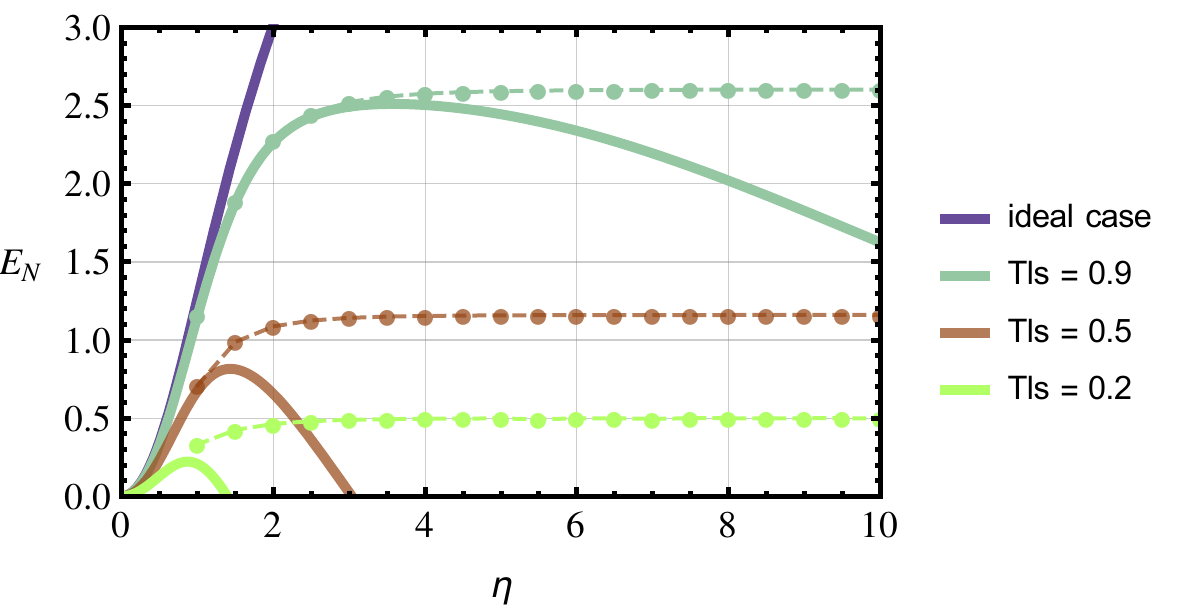}
\caption{\label{aesimple} Logarithmic negativity $E_{N}$ as the function of QND coupling strength $\eta$ in the lossless adiabatic case (top purple curve) and in the case of radiation losses present (lower solid curves). Dotted lines correspond to the case of the optimal combination of the strengths of individual interactions. It demonstrates that radiation losses only partially limit generation of entanglement from the transducer. The optimization of gains is efficient only for large loss.}
\end{figure}

We choose the entanglement of the modes $A$ and $B$ as the measure of the efficiency of the proposed transducer. Our consideration is limited to zero-mean Gaussian states as the initial states of the three modes are such (the vacuum states for the radiation ones and the thermal state for the mechanical mode) and the nondemolition interaction due to its linearity preserves the Gaussianity of the quantum states. Any zero-mean Gaussian state $\hat \rho$ of two modes $A$ and $B$ with quadratures $f = [\mathcal X_A, \mathcal Y_A, \mathcal X_B, \mathcal Y_B]^T$ can be fully described by the covariance matrix with elements $V_{ij} = \frac{1}{2} \Tr \left[ \hat \rho \left( f_i f_j + f_j f_i \right) \right]$. To numerically characterize the Gaussian entanglement we use logarithmic negativity defined as
\begin{equation}
E_{N} = \max \left[ 0, - \log_2 \nu_{-} \right],
\end{equation}
with $\nu_{-}$ being the smallest symplectic eigenvalue of the partially transposed covariance matrix~\cite{weedbrook_gaussian_2012}  that can be computed as follows:
\begin{align*}
\nu_{-} = \frac{1}{\sqrt{2}}\sqrt{\Sigma(V) - \sqrt{\Sigma(V)^2 - 4 \det V}},
\end{align*}
where
\begin{align*}
\Sigma (V) = \det \mathcal V_1 + \det \mathcal V_2 - 2 \det \mathcal V_c,
\end{align*}
with $\mathcal{V}_{1,2}$, $\mathcal V_c$ being $2 \times 2$ block-matrices composing the covariance matrix:
\begin{align*}
V =  \begin{bmatrix}
\mathcal V_1 & \mathcal V_c \\
\mathcal V_c^T & \mathcal V_2
\end{bmatrix}.
\end{align*}
 In the ideal case of adiabatic elimination of the intracavity modes and the absence of decoherence processes the symplectic eigenvalue for the two radiation modes $A$ and $B$ may be expressed in the following form:
\begin{equation}
	\label{eq:logneg_adiabatic}
\nu_-^{0} = \sqrt{1 - 2 \eta^4 \left[ \sqrt{ 1 + \eta^{-4}} - 1 \right]}.
\end{equation}
The corresponding logarithmic negativity is nonzero for arbitrary $\eta >0$ and monotonically increases regardless of the state of the mechanical mediator since in~\eqref{qndAE} the mechanical quadratures appear to be traced out of the transformations of the radiation modes.  The dependence of the entanglement on the coupling strength is illustrated in Fig.~\ref{aesimple} (solid purple curve). The main question is how close the realistic transducer can be to this idealised case.

\section{Decoherence processes for the symmetrical transducer}
\label{decoh}

In this section we consider the radiation loss and mechanical decoherence during whole time of the transducer operation. First we include radiation losses in the delay lines, then we study the influence of the mechanical bath and finally we combine both to explore their joint contribution. It allows a detailed analysis of decoherence in the quantum transducer.

\subsection{The influence of radiation losses}
\label{section_optlosses}
\label{Opt_losses_sect}

The radiation losses may be modeled by a virtual beamsplitter with the transmittance $T_{ls}$. After a mode with quadratures $\mathcal Q$ passes this beamsplitter the quadratures are transformed in the following way:
\begin{equation*}
\mathcal Q \rightarrow \sqrt{T_{ls}} \mathcal Q + \sqrt{1 - T_{ls}} Q_{ls}
\end{equation*}
with $Q_{ls}$ being noise quadratures of vacuum. We introduce these beamsplitters after the first and the second QND interactions (see Fig.~\ref{fig1}~(b)).  For the sake of simplicity we assume damping coefficients $T_{ls}$ to be the same for both modes. We consider the initial mechanical state to be in the ground state within this section.

The radiation losses break the entanglement monotonicity for increased interaction gain $\eta$.  Instead, the maximal value of logarithmic negativity is reached for a finite coupling.  This effect is obviously more pronounced at larger losses (see Fig.~\ref{aesimple} for details).  For higher $\eta$ smaller amounts of losses are sufficient to break the entanglement.
In the limit of small losses $T_{ls} \sim 1$ and weak coupling $\eta \ll 1$ the symplectic eigenvalue may be approximated in the following form:
\begin{equation}
\nu_- \simeq 1 - \frac{1}{2} \left( 1 + T_{ls} \right) \eta^2.
\end{equation}
Losses therefore do not impose a threshold on the value of $\eta$~--- for any transmittance $T_{ls}$ there is entanglement for arbitrarily low values of $\eta$.  The coupling strength $\eta$ however becomes bounded from above (see Fig.~\ref{aesimple}). We also see from this figure that the entanglement behavior near the origin is defined by the losses value~--- the approximated value of the derivative of the logarithmic negativity reads:
\begin{equation}
\frac{\partial E_N}{\partial \eta} \bigg\rvert_{\eta \rightarrow 0} \simeq \frac{2}{\ln 2} \sqrt{T_{ls}} \eta.
\end{equation}

In the ideal case without any losses we compensate for the influence of the noisy mechanical mediator setting strengths of each interaction equal to each other.  Losses lead to the imbalance in the system that can be corrected by making these strengths non-equal.  We numerically find  the optimal combination of individual QND gains $\eta_i$ that provides the maximum of the achievable entanglement given the constraint on coupling strengths $0 < \eta_i < \eta$. The coupling strength can be manipulated by change of interaction time or pumping. They are equivalent at this point. The result of the numerical optimization is presented in Fig.~\ref{aesimple}.  The maximal logarithmic negativity $E_{N}$ is plotted as a function of the upper boundary of the region over which we optimize. As we can see from this figure the optimization helps to restore high values of entanglement especially in case of high losses. In the case of small losses and small $\eta$ non-optimized logarithmic negativity is close to the maximally achievable value~\eqref{eq:logneg_adiabatic}.

\subsection{The influence of the thermal environment}
\label{MB}

\begin{figure*}[!htbp]
	\centering
		\includegraphics[width=.6\linewidth]{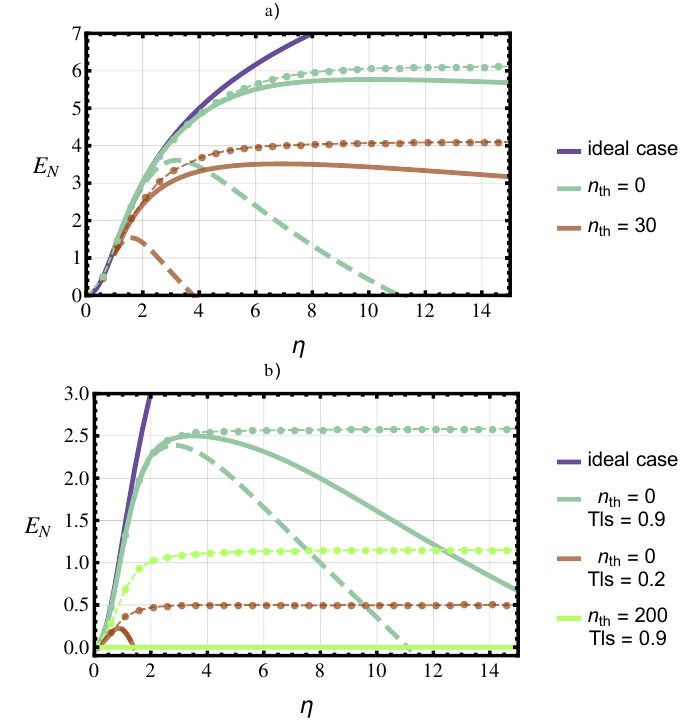}
		\caption{
		Logarithmic negativity $E_N$ as the function of the QND coupling $\eta$. For both left and right figures solid and dashed lines correspond to the non-optimized case with respectively $g$ or $\tau$ varied, dotted lines with markers show the result of optimization. For the plots we used  $\gamma = 1.5 \times 10^{-6} \, \kappa$. (a) Lossless case in presence of mechanical bath. Parameters are varied in the following regions: $0 \leq g \leq 0.4 \kappa$; $ 7 \times 10^2 / \kappa \leq \tau \leq 9 \times 10^4/ \kappa$. This plot demonstrates that mechanical bath does not affect the entanglement drastically and the optimization is efficient for larger mechanical bath occupation numbers. (b)~Mechanical bath and radiation losses. Parameters here are varied in the following regions: $0 \leq g \leq 0.4 \kappa$; $7 \times 10^2 / \kappa \leq \tau \leq 9 \times 10^4/ \kappa$. This plot demonstrates that the performance of the proposed transducer may be quite high. Even for large bath occupation $n_{th} = 200$ the optimization helps to reach significant values of the entanglement.
		}
	\label{fig:final}
\end{figure*}

Now we explore our setup with the presence of the thermal mechanical environment and investigate its influence on the protocol performance. We consider the mechanical bath to be in the thermal state with mean occupation number $n_{th}$, being coupled to the mechanical mode at rate $\gamma$ and we model it by the noisy quadratures $\xi_{xi,pi}$ in~\eqref{QNDLang} with $i=\{1,2 \}$.

In the idealised adiabatic case each of interactions is parametrized by a single coupling parameter~$\eta_i$, upper-bounded by maximal $\eta$. In presence of the mechanical bath the entanglement changes differently with respect to changes in optomechanical coupling $g$ and pulse duration $\tau$ even if those result in equal coupling parameter~$\eta$. It is reflected in Fig.~\ref{fig:final}~(a). Increasing of $g$ causes deviation from the monotonic increase of entanglement which is seen more clearly for large values of $g$.  If we instead increase the temporal duration of pulses $\tau$ to achieve same interaction gain $\eta$ the entanglement is suppressed stronger because the influence of the mechanical bath is obviously more significant for longer interaction times.  It is worth noting that for any thermal occupation arbitrarily low coupling~$\eta$ generates entanglement. We also note that in contrast to previous section, the derivatives near the origin are the same in this case, so in the limit $\eta \ll 1$ all curves coincide.

To reach maximal entanglement we again optimize the logarithmic negativity with respect to the four unequal optomechanical couplings $g_i$ and different interaction times $\tau$. The result of this optimization is presented in Fig.~\ref{fig:final}~(a).  The optimization proves especially useful for larger values of mean occupation number $n_{th}$: in contrast to the non-optimal case of equal couplings, in the optimized regime entanglement monotonically increases with $\eta$.

We would like to note that in the region of small values of the mean bath occupation number $n_{th}$ and QND coupling~$\eta$ the entanglement values are close to the ones of the ideal adiabatic case. On the other hand, in the case of large $n_{th}$, which is of our interest, the entanglement increases at different rate than in the adiabatic regime.

\subsection{Joint influence of radiation losses and mechanical bath}
\label{joint_impact}

 To complete the full analysis we consider the joint impact of the radiation losses and mechanical bath on the protocol performance which is reflected in Fig.~\ref{fig:final}~(b). Apparently, joint influence of the radiation losses and mechanical bath is not critical. The transducer still keeps possibility to generate detectable entanglement, especially for low $\eta$.  The figure  shows that including radiation losses in addition to mechanical bath depresses the curves more for larger interaction strength in agreement with results of Section~\ref{section_optlosses}. The figure shows as well that the influence of the radiation losses is more drastic than the mechanical bath impact when both are present simultaneously. It is therefore important to keep the delay lines lossless.

We once again optimize the interaction parameters $\eta_i$ by varying simultaneously optomechanical couplings $g_i$ and the pulse durations $\tau_i$ to achieve maximum of entanglement.  We see that in the case of joint influence of both mechanical bath and radiation losses the optimization helps sufficiently, especially for larger values of mean occupation number $n_{th}$ and radiation losses. In particular, for high $n_{th} = 200$ where for the non-optimized case the entanglement is not generated at all, the optimal regime shows sufficient values of logarithmic negativity which is a very promising result. In agreement with Section~\ref{Opt_losses_sect} the optimal entanglement saturates to a finite value. It clearly demonstrates advantage of the optimization beyond the basic idea of geometric phase effect in the real pulsed cavity quantum transducer.

\section{Asymmetrical transducer}
\label{asymmetr_sect}

\begin{figure}[htbp!]
\centering \includegraphics[width=0.7\linewidth]{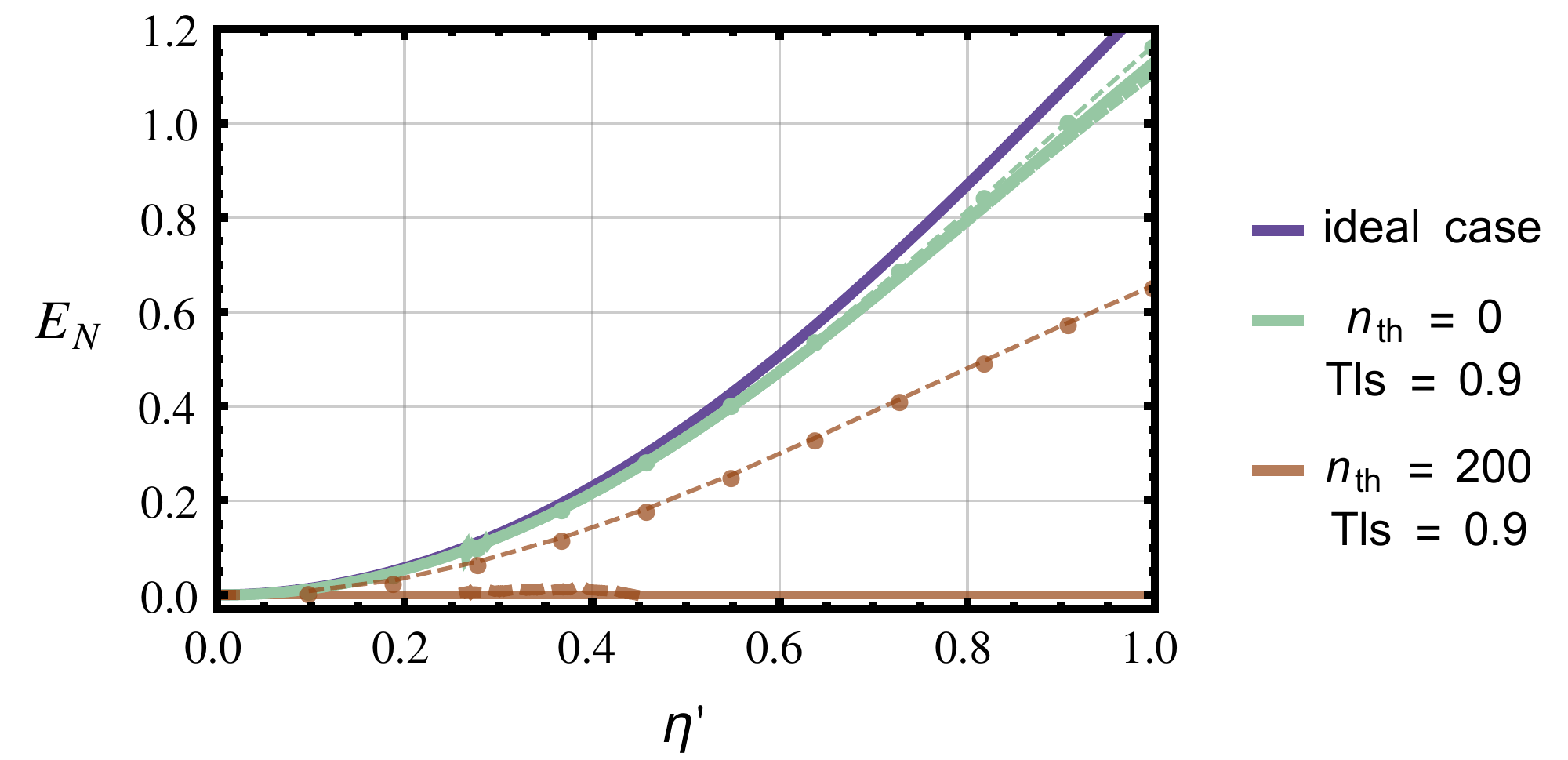}
\caption{\label{asymm_plot} Logarithmic negativity $E_{N}$ as the function of QND coupling $\eta '$ in the presence of mechanical bath and radiation losses for the case of asymmetric transducer. Solid lines correspond to the non-optimized cases with $g_{A,B}$ varied, dashed lines~--- for the same case with $\tau_{A,B}$ increased, dotted lines with markers stand for the optimized cases. Parameters are varied in the following regions: $0 \leq g_A \leq 0.07 \kappa_A$; $0 \leq g_B \leq 0.1 \kappa_B$; $2.2 \times 10^2 /\kappa_A \leq \tau_A \leq 4.4\times 10^3/\kappa_A$; $2.3/\kappa_B \leq \tau_B \leq 113/\kappa_B$ with $\kappa_B = 0.01\times \kappa_A$ and $\gamma = 1.5~\times~10^{-4} \kappa_B $. Brown dashed line is responsible for the changes in $\tau_B$ while changes in $\tau_A$ in the corresponding region does not lead to any entanglement appearance thus demonstrating the asymmetry of the system. This figure demonstrates that the proposed transducer is feasible to entangle optical and microwave fields with the state of the art experimental possibilities.}
\end{figure}

Up to this moment we considered a symmetric transducer that could in principle be implemented with two radiation modes entering in turns a same optomechanical cavity.  Now we switch to a principally different case, where a common mechanical mode is shared across two opto(electro-)mechanical cavities.  This type of transducer allows coupling physically different modes of radiation, for instance, optical and microwave fields and was implemented in continuous wave regime in Refs.~\cite{Bagci2014, Andrews2014}.  The performance of the transducer has been proven~\cite{Andrews2014} to be partially limited by mechanical environment occupation.  We show that our scheme is capable of reasonable performance at relatively high temperatures.

To characterize the system performance we introduce new effective QND coupling parameter $\eta ' = \sqrt{2 g_A g_B \sqrt{\frac{\tau_A \tau_B}{\kappa_A \kappa_B}}}$, where subscripts $A$ and $B$ denote optical and microwave systems correspondingly. We take into account losses in both modes and the mechanical bath.

Our analysis shows that in the case of small radiation losses and low mechanical bath occupation the optoelectromechanical transducer demonstrates very small deviations from maximally achievable performance (See Fig.~\ref{asymm_plot}) even without optimization of the individual interactions.  For the numerical parameters of the analysis we were inspired by two experimental works.  The first one reported in Ref.~\cite{Meenehan2015} considers a nanoscale silicon optomechanical crystal and the second one~\cite{Palomaki2013} explores pulsed entanglement in an electromechanical system.

To achieve maximum of entanglement we performed the same optimization of parameters as we did previously unless this time the coupling strengths $g_{A,B}$ and pulse durations $\tau_{A,B}$ for the two modes were bounded in individual regions in order to reflect the difference between the two modes. In the case of small losses and low occupations this optimization does not help sufficiently (See Fig.~\ref{asymm_plot} where green lines virtually overlap).  However for the case of high thermal occupation $n_{th} = 200$ where the entanglement is not observable (solid brown line on Fig.~\ref{asymm_plot}) the optimized curve demonstrates significant values of the logarithmic negativity (brown dotted line with markers).

It is worth noting that variation of $\tau_A$ and $\tau_B$ leads to different results in the entanglement behavior. As you may see for the case of $n_{th} = 200$ variation of $\tau_B$ (dashed brown curve) leads to a region of non-zero entanglement, whereas variation of $\tau_A$ in the region of realistic parameters does not produce any entanglement. This is related to the fact that at large bath occupation number the individual subsystems are very sensitive to the changes in the pulse duration thus the asymmetry of the transducer becomes more apparent.

\section{Conclusion}
\label{conclusion}

In this paper we explored pulsed optomechanical transducer which entangles two directly non-interacting radiation modes with assistance of a noisy mechanical mediator. We considered this system in the adiabatic regime for the case of symmetrical transducer interconnecting optical fields and we explored the realistic performance of this system in the presence of decoherence for both symmetrical and asymmetrical (allowing to connect optical to microwave radiation) cases. We have shown that the appropriate choice of parameters and their numerical optimization over controllable interaction time and pumping power allow promising performance of such device for experiment. It goes beyond simple understanding based on the geometric phase effect~\cite{Kupcik2015}. Particularly, for the case of very high bath occupation number $n_{th} = 200$ where the entanglement is not generated in the non-optimal case, the optimization shows significant values of achievable logarithmic negativity. It is apparently measurable value of the entanglement detectable in the experiments. This result is a good demonstration of potential efficiency of proposed pulsed transducer and a stimulation for the experimental teams. It opens the way for further exploration of pulsed transducers combining them with other physical platforms, like for example atoms~\cite{Hammerer2009,Hammerer2009a} or solid-state systems like NV centers~\cite{Maurer2012, Huck2011}.

\section*{Funding}

Czech Science Foundation (GAČR) (GB14-36681G); Institutional Support of Palacký University (IGA-PřF-2017-008).



\begin{thebibliography}{99}
\bibitem {Kimble2008} H.~J. Kimble, ``The quantum
  internet,'' Nature {\bfseries 453}, 1023–1030 (2008).
\bibitem{Verdu2009} J.~Verd\'u, H.~Zoubi, Ch.~Koller, J.~Majer, H.~Ritsch, and J.~Schmiedmayer, ``Strong magnetic coupling of an ultracold gas to a superconducting waveguide cavity,'' Phys. Rev. Lett. {\bfseries 103}, 043603 (2009).
\bibitem{Marcos2010} D.~Marcos, M.~Wubs, J.~M. Taylor, R. Aguado, M.~D. Lukin, and A.~S. Sørensen, ``Coupling nitrogen-vacancy centers in diamond to superconducting flux qubits,'' Phys. Rev. Lett. {\bfseries 105}, 210501 (2010).
\bibitem{Hafezi2012} M. Hafezi, Z. Kim, S.~L. Rolston, L.~A. Orozco, B.~L. Lev, and J.~M. Taylor, ``Atomic interface between microwave and optical photos,'' Phys. Rev. A {\bfseries 85}, 020302(R) (2012).
\bibitem{Regal2011} C.~A. Regal and W. Lehnert, ``From cavity electromechanics to cavity optomechanics,'' Journal of Physics: Conference Series {\bfseries 264}, 012025 (2011).
\bibitem{Takeda2013} S. Takeda, T. Mizuta, M. Fuwa, P. van Loock, and A. Furusawa, ``Determenistic quantum teleportation of photonic quantum bits by a hybrid technique'', Nature {\bfseries 500}, 315-318 (2013).
\bibitem{LinTian2015} L. Tian, ``Optoelectromechanical transducer: Reversible conversion between microwave and optical photons,'' Annalen der Physik {\bfseries 527}, 1-14 (2015).
\bibitem{Ogawa2016} H. Ogawa, H. Ohdan, K. Miyata, M. Taguchi, K. Makino, H. Yonezawa, J. Yoshikawa, and A. Furusawa, ``Real-time quadrature measurement of a single-photon wave packet with continuous temporal-mode matching,'' Phys. Rev. Lett. {\bfseries 116}, 233602 (2016).
\bibitem{Makino2016} K. Makino, Y. Hashimoto, J. Yoshikawa, H. Ohdan, T. Toyama, P. van Loock, and A. Furusawa, ``Synchronization of optical photons for quantum information processing,'' Science Advances {\bfseries 2}, e1501772 (2016).
\bibitem{Palomaki2013} T.~A. Palomaki, J.~D. Teufel, R.~W. Simmonds, and K.~W. Lehnert, ``Entangling mechanical motion with microwave fields,'' Science {\bfseries 342}, 710-713 (2013).
\bibitem{Lehnert2013} T.~A. Palomaki, J.~W. Harlow, J.~D. Teufel, R.~W. Simmonds, and K.~W. Lehnert, ``Coherent state transfer between itinerant microwave fields and a mechanical oscillator,'' Nature {\bfseries 495}, 210-214 (2013).
\bibitem{Hofer2011} S.~G. Hofer, W. Wieczorek, M. Aspelmeyer, and K. Hammerer, ``Quantum entanglement and teleportation in pulsed cavity optomechanics,'' Phys. Rev. A. {\bfseries 84}, 052327 (2011).
\bibitem{Berry1987} M.~V. Berry, ``The adiabatic phase and Pancharatnam's phase for polarized light,'' J. Mod. Opt. {\bfseries 34}, 1401-1407 (1987).
\bibitem{Kupcik2015} V. Kup\v{c}\'ik and Radim Filip, ``Continous-variable entanglement mediated by a thermal oscillator,'' Phys. Rev. {\bfseries 92}, 022346 (2015).
\bibitem{Aspelmeyer2014} M. Aspelmeyer, T.~J. Kippenberg, and F. Marquardt, ``Cavity optomechanics,'' Rev. Mod. Phys. {\bfseries 86}, 1391 (2014).
\bibitem{Metcalfe2014} M. Metcalfe, ``Applications of cavity optomechanics,'' Appl. Phys. Rev. {\bfseries 1}, 031105 (2014).
\bibitem{Barzanjeh2012} Sh. Barzanjeh, M. Abdi, G.~J. Milburn, P. Tombesi, and D. Vitali, ``Reversible Optical-to-Microwave Quantum Interface,'' Phys. Rev. Lett. {\bfseries 109}, 130503 (2012).
\bibitem{Wang2012} Y-D Wang and A.~A. Clerk, ``Using interference for high fidelity quantum state transfer in optomechanics,'' Phys. Rev. Lett. {\bfseries 108}, 153604 (2012).
\bibitem{Lin2012} L. Tian, ``Adiabatic state conversion and pulse transmission in optomechanical system,'' Phys. Rev. Lett. {\bfseries 108}, 153604 (2012).
\bibitem{YDWang2012} Y-D Wang and A.~A. Clerk, ``Using dark modes for high-fidelity optomechanical quantum state transfer,'' New. J. Phys. {\bfseries 14}, 105010 (2012).
\bibitem{McGee2013} S.~A. McGee, D. Meiser, C.~A. Regal, K.~W. Lehnert, and M.~J. Holland, ``Mechanical resonators for storage and transfer of electrical and optical quantum states,'' Phys. Rev. A {\bfseries 87}, 053818 (2013).
\bibitem{Winger2011} M. Winger, T.~D. Blasius, T.~P. Mayer Alegre, A.~H. Safavi-Naeini, S. Meenehan, J.~Cohen, S.~Stobbe, and O. Painted, ``A chip-scale integrated cavity-electro-optomechanics platform,'' Optics Express {\bfseries 19}, 24905-24921 (2011)
\bibitem{Bochmann2013} J. Bochmann, A. Vainsencher, D.~D. Awschalom, and A.~N. Cleland, ``Nanomechanical coupling between microwave and optical photons,'' Nature Physics {\bfseries 9}, 712-716 (2013).
\bibitem{Andrews2014} R.~W. Andrews, R.~W. Peterson, T.~P. Purdy, K. Cicak, R.~W. Simmonds, C.~A. Regal, and K.~W. Lehnert, ``Bidirectional and efficient conversion between microwave and optical light,'' Nature Physics {\bfseries 10}, 321-326 (2014).
\bibitem{Bagci2014} T. Bagci, A. Simonsen, S. Schmid, L.~G. Villanueva, E. Zeuthen, J. Appel, J.M. Taylor, A.~S. Sørensen, K. Usami, A. Schliesser, and  E.~S. Polzik, ``Optical detection of radio waves through a nanomechanical transducer,'' Nature {\bfseries 507}, 81-85 (2014).
\bibitem{Menke2017} T. Menke, P.~S. Burns, A.~P. Higginbotham, N.~S. Kampel, R.~W. Peterson, K. Cicak, R.~W. Simmonds, C.~A. Regal, and K.~W. Lehnert, ``Reconfigurable re-entrant cavity for wireless coupling to an electro-optomechanical device,'' https://arxiv.org/abs/1703.06470.
\bibitem{Asjad2015} M. Asjad, S. Zippilli, P. Tombesi, and D. Vitali, ``Large distance continuous variable communication with concatenated swaps,'' Physica Scripta, {\bfseries 90}, 7 (2015).
\bibitem{Asjad2016} M. Asjad, P. Tombesi, and D. Vitali, ``Feedback control of two-mode output entanglement and steering in cavity optomechanics,'' Phys. Rev. A {\bfseries 94}, 052312 (2016).
\bibitem{Tian2010} L. Tian and H. Wang, ``Optical wavelength conversion of quantum states with optomechanics,'' Phys. Rev. A. {\bfseries 82}, 053806 (2010).
\bibitem{Dong2015} C. Dong, V. Fiore, M.~C. Kuzyk, L. Tian, and H. Wang, ``Optical wavelength conversion via optomechanical coupling in a silica resonator,'' Ann. Phys. (Berlin) {\bfseries 527}, 100 (2015).
\bibitem{Miwa2014} Y. Miwa, J.-I. Yoshikawa, N. Iwata, M. Endo, P. Marek, R. Filip, P. van Loock, and A. Furusawa, ``Exploring a New Regime for Processing Optical Qubits: Squeezing and Unsqueezing Single Photons,'' Phys. Rev. Lett. {\bfseries 113}, 013601 (2014).
\bibitem{Aspelmeyer_book2014} P. Treutlein, C. Genes, K. Hammerer, M. Poggio, and P. Rabl, { \it Cavity optomechanics}, (Springer, Berlin Heidelberg, 2014) Chap. Hybrid Mechanical Systems.
\bibitem{Wallquist2009} M. Wallquist, K. Hammerer, P. Rabl, M. Lukin, and P. Zoller, ``Hybrid quantum devices and quantum engineering,'' Phys. Scr. {\bfseries 2009}, 014001 (2009).
\bibitem{braginsky_quantum_1980} V.~B. Braginsky, Y.~I. Vorontsov, and K.~S. Thorne, ``Quantum Nondemolition Measurements,'' Science {\bfseries 209}, 547-557 (1980).
\bibitem{clerk_back-action_2008} A.~A. Clerk, F. Marquardt, and K. Jacobs, ``Back-action evasion ans squeezing of a mechanical resonator using a cavity detector,'' New J. Phys. {\bfseries 10}, 095010 (2008).
\bibitem{Sorensen2000} A.~S. Sørensen and K. Mølmer, ``Entanglement and quantum computation with ions in thermal motion,'' Phys. Rev. A. {\bfseries 62}, 022311 (2000).
\bibitem{Milburn2000} G.~J. Milburn, S. Schneider, and D.~F.~V. James, ``Ion trap quantum computing with warm atoms,'' Fortschr. Phys. {\bfseries 48}, 801-10 (2000).
\bibitem{Leibfried2003} D. Leibfried, B. DeMarco, V. Meyer, D. Lucas, M. Barett, J. Britton, W. M. Itano, B. Jelenkovi\'c, C. Lange, T. Rosenband, and D.~J. Wineland, ``Experimental demonstration of a robust high-fidelity geometric two ion-qubit phase gate,'' Nature {\bfseries 422}, 412-5 (2003)
\bibitem{Vacanti2012} G. Vacanti, R. Fazio, M.~S. Kim, G.~M. Palme, M. Paternostro, and V. Vedral, ``Geometric phase kickback in a mesoscopic qubit-oscillator system,'' Phys. Rev. A {\bfseries 85}, 022129 (2012).
\bibitem{Pikovsky2012} I. Pikovsky, M.~R. Vanner, M. Aspelmeyer, M.~S. Kim, and \v{C}. Brukner, ``Probing Planck-scale physics with quantum optics,'' Nature Phys. {\bfseries 8}, 393-7 (2012).
\bibitem{Khosla2013} K.~E. Khosla, M.~R. Vanner, W.~P. Bowen, and G.~J. Milburn, ``Quantum state preparation of a mechanical resonator using an optomechanical geometric phase,'' New. J. Phys. {\bfseries 15}, 043025 (2013).
\bibitem{law_interaction_1995} C.~K. Law, ``Interaction between a moving mirror and radiation pressure: A Hamiltonian formulation,'' Phys. Rev. A {\bfseries 51}, 2537-2541 (1995).
\bibitem{Giovannetti2001} V. Giovannetti and D. Vitali, ``Phase-noise measurement in a cavity with a movable mirror undergoing quantum Brownian motion,'' Phys. Rev. A {\bfseries 63}, 023812 (2001).
\bibitem{rakhubovsky_squeezer-based_2016} A.~A. Rakhubovsky, N. Vostrosablin, and R. Filip, ``Squeezer-based pulsed optomechanical interface,'' Phys. Rev. A {\bfseries 93}, 033813 (2016).
\bibitem{weedbrook_gaussian_2012} C. Weedbrook, S. Pirandola, R. Garc\'ia-Patr\'on, N.~J. Cerf, T.~C. Ralph, J.~H. Shapiro, and S. Lloyd, ``Gaussian quantum information,'' Rev. Mod. Phys. {\bfseries 84}, 621-669 (2012).
\bibitem{Meenehan2015} S.~M. Meenehan, J.~D. Cohen, G.~S. MacCabe, F. Marsili, M.~D. Shaw, and O. Painter, ``Pulsed excitation dynamics of an optomechanical crystal resonator near its quantum ground state of motion,'' Phys. Rev. X {\bfseries 5}, 041002 (2015).
\bibitem{Hammerer2009} K. Hammerer, M. Aspelmeyer, E. S. Polzik, and P. Zoller, ``Establishing Einstein-Poldosky-Rosen Channels between Nanomechanics and Atomic Ensembles,'' Phys. Rev. Lett. {\bfseries 102}, 020501 (2009).
\bibitem{Hammerer2009a} K. Hammerer, M. Wallquist, C. Genes, M. Ludwig, F. Marquardt, P. Treutlein, P. Zoller, J. Ye, and H. J. Kimble, ``Strong Coupling of a Mechanical Oscillator and a Single Atom,'' Phys. Rev. Lett. {\bfseries 103}, 063005 (2009).
\bibitem{Maurer2012} P. C. Maurer, G. Kucsko, C. Latta, L. Jiang, N.Y. Yao, S. D. Bennett, F. Pastawski, D. Hunger, N. Chisholm, M. Markham, D. J. Twitchen, J. I. Cirac, and M. D. Lukin, ``Room-Temperature Quantum Bit Memory Exceeding One Second,'' Science, {\bfseries 336}, 1283-1286 (2012).
\bibitem{Huck2011} A. Huck, S. Kumar, A. Shakoor, and U. L. Andersen, ``Controlled Coupling of a Single Nitrogen-Vacancy Center to a Silver Nanowire,'' Phys. Rev. Lett., {\bfseries 106}, 096801 (2011).

\end{thebibliography}
\end{document}